\documentclass[twocolumn,showpacs,amsmath,amssymb]{revtex4}
\usepackage{graphicx}
\usepackage{dcolumn}
\usepackage{bm}
\newcommand{\be}{\begin{eqnarray}}
\newcommand{\en}{\end{eqnarray}}
\newcommand{\ben}{\begin{eqnarray}}
\newcommand{\enn}{\end{eqnarray}}

\newcommand{\bi}{\begin{itemize}}
\newcommand{\ei}{\end{itemize}}

\newcommand{\la}{\langle}
\newcommand{\ra}{\rangle}

\def\jsm#1#2#3#{J.Stat.Mech. {\bf#1}, #2 (#3)}
\begin{document}
\title{Work probability distribution and tossing a biased coin}
\author{Arnab Saha}\email{arnab@bose.res.in}
\author{Jayanta K Bhattacharjee}\email{jkb@bose.res.in}
\affiliation{S.N. Bose National Centre for Basic Sciences, Saltlake, Kolkata 700098, India}
\author{Sagar Chakraborty}\email{sagar@nbi.dk}
\affiliation{NBIA, Niels Bohr Institute, Blegdamsvej 17, 2100 Copenhagen $\O$, Denmark}
\date{\today}
\begin{abstract}
We show that the rare events present in dissipated work that enters Jarzynski equality, when mapped appropriately to the phenomenon of large deviations found in a biased coin toss, are enough to yield a quantitative work probability distribution for Jarzynski equality. This allows us to propose a recipe for constructing work probability distribution independent of the details of any relevant system. The underlying framework, developed herein, is expected to be of use in modelling other physical phenomena where rare events play an important role.
\end{abstract}
\pacs{05.70.ln, 05.90.+m, 05.20.-y}
\maketitle
\section{introduction}
Large deviations play a significant role in non-equilibrium statistical physics\cite{OO,KU,DE,PEL,UDO,PA}. They are difficult to handle because their effects though small, are not amenable to perturbation theory. All the conventional perturbation theories in statistical physics are fashioned about a Gaussian distribution, which almost by definition, is the distribution with no large deviations. This can be seen in static critical phenomena, critical dynamics, dynamics of interfacial growth, statistics of polymer chain and myriad other problems\cite{MM}. Our contention is: in the large deviation theory \cite{20, 21, DEM, 22}, the central role is played by the distribution associated with tossing of a coin and the simple coin toss is the ``Gaussian model" of problems where rare events play significant role. In this paper, we illustrate our contention by applying it to the study of some aspects of Jarzynski equality.
\\
Fluctuation theorems form a very important part of nonequilibrium statistical mechanics \cite{2,3,4,5,6,7,8,9}. There has been a lot of activity in the last decade or so and various forms of such theorems has been established. One particular form is Jarzynski equality \cite{4,5}. If $W$ is the work done during a period of duration $\tau$, during which an external force acts on the system and does work, then Jarzynski established that in units of $K_BT$
\be
\la e^{-W}\ra=e^{-\Delta F}
\label{1}
\en
where angular brackets denote ensemble average and $\Delta F$ is the free energy difference for the equilibrium free energies corresponding to the initial and final states. Here $K_B$ is Boltzmann constant and $T$ is the temperature of the concerned system in the initial equilibrium state or, equivalently, the temperature of the heat reservoir with which the system was thermalized before the process took place. It is important to note that here $W$ is a path (trajectory followed by the system during $\tau$) dependent variable. So, if we consider the ensemble of all possible paths (each path originating from one of the several microstates corresponding to the initial equilibrium macrostate), different values of $W$ along different path can be identified with a set of random variable. Now, if we define another random variable --- dissipative work along a path --- as
\be
W_D\equiv W-\Delta F
\label{DeltaF}
\en
Jarzynski equality (\ref{1}) shows that
\be
\la e^{-W_D}\ra=1
\label{3}
\en
Clearly, to satisfy above equality, $W_D$ should take both positive and negative values. Again, we know that since $\la W\ra$ is the thermodynamic work done in going from initial state to the final state, the second law of thermodynamics would assert that $ \la W\ra \geq \Delta F$ ({\it i.e.}, $\la W_D\ra \geq 0)$, with the equality holding for the reversible process where the system remains always in the equilibrium with its surrounding. So, negative values of $W_D$ are relatively rare events, yet important enough to make the equality of Eq.(\ref{3}) to hold.
\section{The strategy}
The strategy for demonstrating the validity of our approach will be as follows. As the paradigm for the distribution of rare events we will take, as mentioned earlier, the distribution associated with tossing a coin. The random variable associated with a coin toss can range between two finite numbers which we will take to be $0$ and $1$. (It'll be explained in details in the next section.) The mean value $p$, of the variable for an unbiased coin is ${1}/{2}$ and for a biased coin it lies between $0$ and $1$ but $p \neq {1}/{2}$.
Dissipative work along a path $W_D$ ranges from $-\infty$ to $+\infty$ and we will first carry out a transformation that maps it onto the range $0$ to $1$. Further, according to the second law of thermodynamics, the events corresponding to $W > \Delta F$ (or, $W_D > 0$) are more likely than the ones corresponding to $W < \Delta F$ (or, $W_D < 0$) and hence there will be an asymmetry or {\it bias} about the events corresponding to $W=\Delta F$ (or, $W_D=0$). The amount of bias in the statistics of $W$ due to irreversibility is clearly: $\la W \ra - \Delta F$. The corresponding coin has to be biased as well and hence we shall take the asymmetric situation of $p \neq {1}/{2}$. Needless to say that for reversible process, since $\la W\ra=\Delta F$, this bias is zero.
Having defined the mapping --- which, of course, is not unique --- the first check would be to verify if Jarzynski equality in the form of (\ref{3}) is satisfied. This can be tested since we have an explicit distribution --- namely, the one associated with the biased coin toss. Such a check is depicted in figure (1), details regarding which follows later in this paper. From the distribution for $W_D$ one can also have the distribution for $W$. The distributions of work and dissipative work have drawn a lot of attraction.
For various systems these distributions are now known from experiments as well as numerical studies ({\it e.g.} \cite{Imp-Peliti, Douarche, dhar2, dhar3, Lip, BLI, Arnab,dhar1}). Here we obtain these distributions from very general requirements, which are independent of the dynamics (that usually varies from system to system) followed by the systems. Hence it is important to ask whether the experiments on the systems obeying widely different dynamics really exhibit similar distributions. We find the answer here by comparing our results with actual experimental results and numerical simulations. The probability distribution for $W_D$, $P(W_D)$, was obtained experimentally by Liphardt {\it et al.} \cite {Lip}; and, $P(W)$ has been obtained by Blickle {\it et al.} \cite {BLI} for a different system. We have also used an anharmonic oscillator driven by a linear time dependent force to simulate the dynamics and numerically construct $P(W)$ \cite {Arnab}. A similar system has also been studied in \cite{dhar1}.
\\

 In this work, we calculate $P(W)$ from the biased coin toss distribution which we have taken as a starting ansatz based on the principle of large deviations. The only connection between the experiments and the method we use, is the fact that, the experiment is carried out far from equilibrium and hence must feature negative values of $W_D$ (large deviations) and we have started with a distribution which has large deviations built in it. As will turn out, our method will have two parameters which we fix by comparing with the distribution obtained from the experiment. The appropriateness of $P(W)$, we calculate employing the theory of large deviations, is borne out by comparison as well be demonstrated below. The point we want to stress here is that the present theory, which explicitly takes care of large deviations, does not require explicit knowledge of dynamics. Consequently, it has wide range of applicability. The parameters of the distribution need to be fixed in each case from the scales of measured distribution --- {\it e.g.}, peak position and peak magnitude.
\section{coin toss}
We begin by recalling the situation of coin tossing experiment. If we assign a value $1$ to the outcome `heads' and $0$ to the outcome `tails', then the mean after $N$ trials is
\be
M_N=\frac{1}{N}\sum_{i}^N X_i
\label{mn}
\en
This is an experimental mean which belongs to a set of independent and identically distributed ({\it i.i.d}) random numbers lying between $0$ and $1$ since each individual $X_i$ is either $0$ or $1$. For an unbiased coin where ``heads" and ``tails" are equally probable, this mean goes towards ${1}/{2}$ as $N\rightarrow \infty$. However, for a biased coin where the probability of obtaining ``heads" is $p (\neq {1}/{2})$, the mean will converge towards $p$ as $N\rightarrow \infty$. If in $N$ trials `heads' appear $X$ times, then the probability of finding a mean $M_N={X}/{N}$ is
\be
P\left(\frac{X}{N}\right)={}^NC_X p^X(1-p)^{(N-X)}
\label{bin}
\en
This is the binomial distribution, which is the most commonly used example of a theory exhibiting large deviations {\it i.e.} even when $N\gg 1$, we find a $P$ which falls off slower than Gaussian.
$P\left(X/N\right) \sim \exp[-{N\left({X}/{N}-p\right)^2}/{\sigma^2}]$. Taking log of both side of eq. (\ref{bin}), we arrive at:
\ben
\ln P\left(\frac{X}{N}\right)&=&X\ln p+(N-X)\ln (1-p)+ \nonumber\\
&&\ln N!-\ln X!-\ln (N-X)!.
\label{logg}
\enn
We know from Stirling's formula that for large $N$, $N!\simeq N^N e^{-N}\sqrt{2\pi N}$. Applying Stirling's approximation
for large $N$, $X$, $(N-X)$, the above relation becomes
\be
\ln P(x)=-NJ(x),
\label{stir1}
\en
where,
\ben
J(x)&\equiv&x\ln \frac{x}{p}+(1-x)\ln \left(\frac{1-x}{1-p}\right)+ \nonumber \\
&&\frac{1}{2N}\ln x(1-x)+\frac{1}{2N}\ln N
\label{stir2}
\enn
and $x\equiv {X}/{N}$. This leads to
\be
P(x)&\sim& \frac{1}{\sqrt{Nx(1-x)}}\exp(-NI(x))
\en
where,
\be
I(x)\equiv x\ln\frac{x}{p}+(1-x)\ln\frac{1-x}{1-p}
\label{stir3}
\en
For large enough $N$ the pre-factor changes slowly compared to the exponential term and we get Chernoff's formula where $I(x)$ is the rate function.
\section{The mapping}
In order to appreciate the analogy between the large deviation theory and the Jarzynski equality, we need to consider evolution of a relevant system to have a stochastic component, {\it i.e.} there is a regular time dependent force acting on the system from $t=0$ to $t=\tau$ as it proceeds from an initial equilibrium state to a final state and in addition there is a random component. Langevin dynamics of a particle of mass $m$ can be considered as a simple example of such a stochastic dynamics:
\be
m\ddot x = -\frac{\partial V}{\partial x}-\lambda \dot x +f(t)+\eta(t)
\en
where $V$ is a potential function, $f(t)$ is a regular time dependent force, $\lambda$ is damping coefficient and $\eta(t)$ is random noise. Here dots represent time derivative. Fluctuation theorems were proven for such a system by Kurchan \cite{kur} and subsequently by several authors \cite{leb,farago,cohen,dhar}. The time dependent force is switched on at $t=0$ and switched off at $t=\tau$. At $t=0$ the system resides in a macroscopic equilibrium state, corresponding to which there exist a large number of microstates.
Since we are considering stochastic evolution, we can start from the {\it same microstate} and do the experiment $N$ times, each time getting different value of the dissipative work. If $w_D^{i}$ is the dissipative work for $i^{th}$ realisation then we can define $W_D$ [the analogue of $M_N$ is Eq.(\ref{mn})] as
\be
W_D=\frac{1}{N}\sum_{i=1}^Nw_D^i.
\label{wd}
\en
The distribution of $W_D$ is sought from the large deviation principle.
\par
 We now note that $\la W \ra \geq \Delta F$ (or, $\la W_D\ra \geq 0$) according to the second law of thermodynamics --- the equality holds for reversible processes. To implement our scheme, we need to define a transformation which maps $W_D$ to another variable $Z$ such that, $0\leq Z \leq 1$, in accordance with the experimental mean in coin toss scenario, given in Eq.(\ref{mn}). We consider the variable $W_D+c$, where `$c$' is a quantity that we shall fix later. The class of transformation we consider here is
\be
Z(W_D)=\frac{1}{2}[1-\tanh \alpha(W_D+c)],
\label{trans}
\en
where `$\alpha$' is a parameter which eventually will have to be fixed using experimental results. Actually, only the positive constant `$\alpha$' defines this class of transformations because a constraining relation for $c$ will be established. Our ansatz is that,  $Z$, like $W_D$, satisfies large deviation principle and the rate function for the coin toss problem is the rate function for $Z$. So, the rate function for $Z$ is
\be
I(Z)=Z\ln\frac{Z}{p}+(1-Z)\ln\frac{1-Z}{1-p}.
\label{6}
\en
The probability distribution for $Z$ is simply
\be
P(Z)\sim \frac{1}{\sqrt{NZ(1-Z)}}e^{-NI(Z)}
\label{7}
\en
where $N$ is the number of trajectories used in constructing the experimental mean of the {\it i.i.d} variables. In our case, $X_i$ in coin tossing experiment and dissipative work  $w_D^i$ in Eq.(\ref{wd}) are {\it i.i.d} variables. We note that $I(Z=1)=\ln({1}/{p})$, while $I(Z=0)=\ln[{1}/({1-p})]$. For $p < {1}/{2}$, $I(Z=0) < I(Z=1)$. The function $I(Z)$ has a minimum at $Z=p$. Thus the probability $P(Z)$ has a peak at $Z=p$ and is exponentially small at $Z=0$ and $Z=1$, but with $P(Z=0)> P(Z=1)$, because of the inequality in $I(Z)$. From second law of thermodynamics, we need $\la W_D\ra > 0$ for irreversible process, {\it i.e.} realizations with the outcome $W_D>0$ is more probable than that of $W_D<0$. All the above constraints are met since $Z\rightarrow 0$ as $W_D\rightarrow \infty$ and $Z\rightarrow 1$ as $W_D\rightarrow -\infty$.
We now return to Eq.(\ref{trans}); noting that $e^{-W_D}=e^c {[{Z}/{(1-Z)}]}^{{1}/{2\alpha}}$, we have
\be
\la e^{-W_D}\ra=\frac{e^c\int_0^1 {\left(\frac{Z}{1-Z}\right)}^{\frac{1}{2\alpha}}P(Z)dZ}{\int_0^1 P(Z)dZ}
\label{m4}
\en
The right hand side of equation (\ref{m4}) is plotted as a function of $N$ in figure (1) for different values of $\alpha$. As $N \rightarrow \infty$, we find $\la e^{-W_D}\ra$ converges to unity for all $\alpha$, as it should according to Jarzynski equality. From figure (1) one can see that as $\alpha$ decreases, lesser number of trajectories (or, realizations) are required for the convergence. The convergence is verified for various values of $p$. The result shown in figure (1) is for $p=0.25$. 

\begin{figure}
\includegraphics[width=0.40\textwidth]{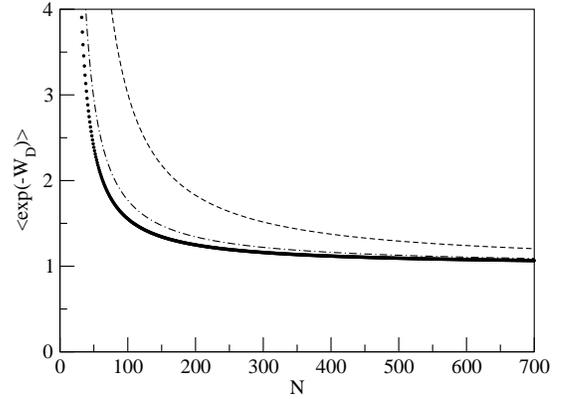}
\caption{In this figure we show the convergence of $\la \exp(-W_D)\ra$ with respect to $N$ for different values of $\alpha$. The thick-dotted, dot-dashed, and dashed lines are respectively for $\alpha=0.17, 0.57, 0.96$. The curves in the figure are obtained by numerically integrating equation (\ref {m4}) by employing Simpson's one-third rule.}
\end{figure}

\section{comparision with experiments}
We will now obtain $P(W)$ from $P(W_D)$ following the theoretical technique discussed above and compare with the work distribution function obtained experimentally and numerically. From eq. (\ref{trans}) we write
\be
W = \Delta F -c +\frac{1}{2\alpha}\ln\frac{1-Z}{Z}
\label{work}
\en
We fix $c=\Delta F$ to get the following simple form
\be
W = \frac{1}{2\alpha}\ln\frac{1-Z}{Z}
\label{work1}
\en
We can now find $P(W)$ by noting the normalisation condition: \be\int_0^1 P(Z)dZ = \int_{-\infty}^{\infty}P(Z=f(W))\left|\frac{dZ}{dW}\right|dW=1\en where, $P(Z)$ is given in Eq.(\ref{7}). The work distribution is then found as
\be
P(W)= P(Z=f(W))\left|\frac {dZ}{dW}\right|
\label{ldp-distri}
\en
with $f(W)=(1-\tanh\alpha W)/2$ and $P(Z)$ as given by Eq.(\ref{7}). The parameter $N$ in Eq.(\ref{7}) can be linked to the width $\sigma$ of the distribution by appealing to the Gaussian limit which shows that $\sigma^2={2p(1-p)}/{N}$. In principle we need to fix three unknown parameters in Eq.(\ref{ldp-distri}), {\it viz.}, $\sigma$, $\alpha$ and $p$. To reduce the task of parameter-adjustments, we pre-assign a value of $p$. As a result, only $\alpha$ and $\sigma$ will be used as fitting parameters.
\\
We now show the comparison between our assertion of the form of $P(W_D)$ [and hence $P(W)$] and different experimental and numerical results.
\subsection{Experiment by Liphardt {\it et al.}} This experiment tests Jarzynski equality by streching a single RNA molecule between two conformations --- both reversibly and irreversibly. The experiment has been done for three different molecular end-to-end extensions and for each extension, three different streching rates are considered. For all combinations of extensions and pulling rates, the experiment provides $P(W_D)$. For the present purpose, we consider three distributions corresponding to 15 nm extension, which are shown in figure (2a). Other distributions can also be taken care of similarly. In this work, $P(W_D)$ we compute, depends on two parameters {\it viz.}, $\alpha$ and $\sigma$. We determine these two parameters by comparing with $P(0)$ and $P_{max}(W_D)$ [maximum value of $P(W_D)$] of the corresponding distribution given by Liphardt {\it et al.}. After fixing these two parameters as $(\alpha, \sigma) \simeq (0.12, 0.07), (0.12, 0.13), (0.14, 0.20)$ for three different pulling rates, we arrive at the full distributions for every pulling rate. This is shown in figure (2b). We fix $p=0.48$ here.
\subsection{ Experiment by Blickle {\it et al.}} This experiment deals with the thermodynamics of an overdamped colloidal particle in a time dependent nonharmonic potential. Blickle {\it et al.} have not only measured $P(W)$, they have also computed $P(W)$ from the relevant Fokker-Planck dynamics. Our explanation of their $P(W)$ is dependent on the choice of the two parameters $\alpha$ and $\sigma$. We determine these parameters by comparing $P(0)$ and $P_{max}(W)$. This fixes $\alpha \simeq 0.1$ and $\sigma \simeq 0.2$. This comparison is shown in figure (3a) and (3b). We fix $p=0.24$ here. The moments found by Blickle {\it et al.} and us compare as follows (Table I)

\begin{center}
\begin{tabular}{|c|c|c|}\hline Moments & Values from experiment& Values from the \\ &  by Blickle {\it et al.}& theory presented here \\ \hline $\la W\ra$ & 2.4 & 2.40 \\\hline $\la W^2\ra$ & 11.6 & 11.74 \\ \hline $\la W^3\ra$ & 63.7 & 64.11 \\ \hline
\end{tabular}
\end{center}

\begin{figure}
\vskip 0.50cm
\hskip 0.10cm
\includegraphics[width=8cm,height=6cm]{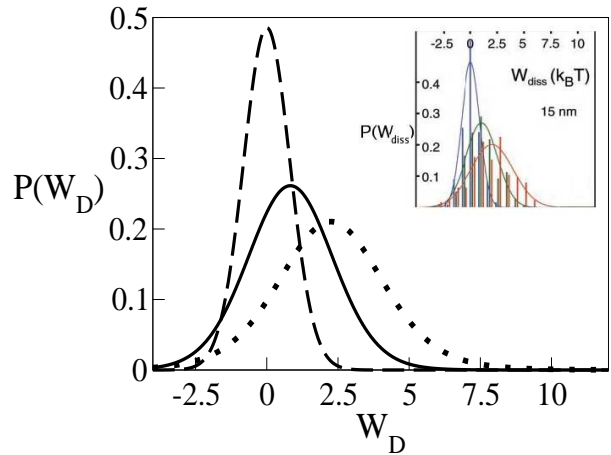}
\caption{({\it Colour online}) The figure of the inset is taken from \cite {Lip} where dissipated work probability distributions are experimentally obtained for a particular end-to-end extension ($= 15 nm$) of P5abc RNA molecule but for three different pulling rates indicated by three different curves --- dashed, solid and dotted. In main figure we obtain $P(W_D)$ by fixing $\alpha$ and $\sigma$, as required by the theory presented here.}
\end{figure}

\begin{figure}
\includegraphics[width=8cm,height=6cm]{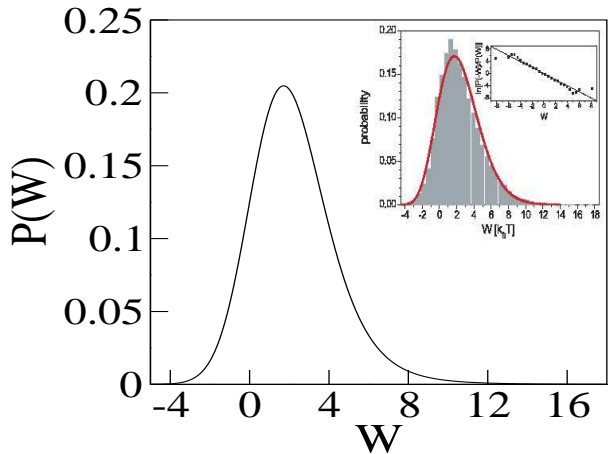}
\caption{({\it Colour online}) The figure of the inset is taken from \cite{BLI}, which shows experimentally as well as numerically obtained work probability distribution function for an overdamped colloidal particle in a time-dependent nonharmonic potential. In the main figure we obtain $P(W)$ from the theory presented here.}
\end{figure}

Since we use only two parameters ($\alpha$ and $\sigma$) to get the distribution, it implies that only $\la W\ra$ and $\la W^2\ra$ have been used. This leaves the $\la W^3\ra$ as a prediction which can be compared with the experimental data. A more sensitive quantity to measure asymmetry of a distribution is $\la \Delta W^3\ra$, where $\Delta W=W-\la W\ra$. Our distribution shows that this moment is nonzero. If $p\neq {1}/{2}$, then ${|\la \Delta W^3\ra|}/({\sigma |1-2p|\la \Delta W^2\ra^{3/2})}$ is a constant, the value of which in our case is 10. Nonzero $\la \Delta W^3\ra$ corresponds to asymmetry of $P(W)$ that has been observed whenever the dynamics has been nonlinear\cite{Arnab,dhar1}. In those cases the cause of the asymmetry is the strength of the nonlinear term. Here the role is played by $(1-2p)$ (though, no particular dynamics is explicitly involved here) and it can be considered as a measure of asymmetry.
\subsection{Driven anharmonic oscillator} We consider here a Brownian particle, trapped by the potential $V(x)= kx^2+\gamma x^4$ (where $k$ and $\gamma$ are constants) and driven by a linearly time-dependent force $f(t)$. The evolution is taken to be governed by following overdamped Langevin dynamics,
\be
\lambda \dot x + \frac{\partial V}{\partial x} = f(t)+\eta(t).
\label{anharmonic}
\en
Here $\eta(t)$ is the random noise coming from heat bath. We assume $\la \eta(t)\ra = 0$ and $\la \eta(t)\eta(t^{\prime})\ra=2T\lambda \delta (t-t^{\prime})$, where $T$ is temperature of the bath. The force $f(t)$ acts from $t=0$ to $t=\tau$ and $P(W)$ (where $W=-\int_0^{\tau}\dot f(t)x(t)dt$) is numerically obtained. The comparison between numerically obtained $P(W)$ and that obtained from Eq.(\ref {ldp-distri}) with $\alpha \simeq 0.15$ and $\sigma \simeq 0.062$ is shown in figure (4). We fix $p=0.28$ here.

\begin{figure}
\vskip 0.50cm
\hskip 0.10cm
\includegraphics[width=8cm,height=6cm]{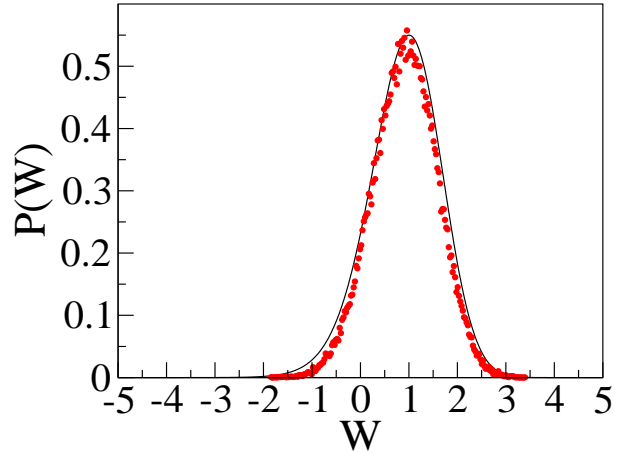}
\caption{({\it Colour online}) In this figure we show how $P(W)$, calculated by simulating the dynamics of a driven Brownian particle obeying Eq.(\ref{anharmonic}) (shown in red dots) collapses to the $P(W)$ calculated from the distribution for tossing a biased coin (shown in black solid line), as it is prescribed here. For simulating the dynamics we take $k=1, T=1, \lambda=1$ and $\gamma=0.1$.}
\end{figure}
\section{Conclusions}

In conclusion, we re-stress that the discussed framework is very general and simple because we require only few parameters from experiments and elementary results from large deviation theory to construct a full work probability distribution, bypassing the `nitty-gritty' of dynamics. We believe that it is possible to make better contact with experiments by constructing more appropriate form for the function $Z$.
Readers would also appreciate that we could derive results concerning Jarzynski equality merely by focusing on the rare events --- rare negative dissipation --- that enter into Jarzynski equality and mapping them onto the biased coin-toss-experiments.
Here we have discussed situations where the evolution had a stochastic component in addition to the regular time dependent force. We hope to extend it to the deterministic situations. In the deterministic case, we envisage the following picture. The evolution of a nonlinear system under time dependent drive is intrinsically chaotic and we can exploit that to define an ``experimental mean" for $w^i_D$. In this case, we need to consider the different initial conditions around an $\epsilon$-neighbourhood $(\epsilon \rightarrow 0)$ of a given microstate and since the evolution of each initial microstate (from same initial macrostate) will be different from each other due to the chaotic flow, we can define $W_D$ as in Eq.(\ref{wd}).
Therefore, in accordance with our contention that the simple coin toss is the `Gaussian model' for the problems where rare events play significant role, one might speculate that the phenomenon of intermittency (and hence multifractality) in fluid turbulence can be obtained by treating rare events in the energy dissipation rate in the similar fashion outlined in this paper. As we have reported elsewhere\cite{arXiv} that, for fluid turbulence, the rare events present in the distribution of energy in the real space, when mapped appropriately on the phenomenon of large deviations found in simple coin toss are enough to yield anomalous exponents which are known to be the signatures of multifractality in fluid turbulence. Within this very framework, we hope to model various other physical phenomena where rare events play a significant role; after all, now we have a working approach to arrive at quantitative results for such processes that cannot be usually solved otherwise.
\acknowledgements
The authors gratefully acknowledge the academic and the financial supports from WBUT (Kolkata) and SNBNCBS (Kolkata). SC  thanks NBIA (Copenhagen) and Danish Research Council for an FNU grant for such supports. AS and JKB thank Prof. Dr. U. Seifert for useful comments and discussions.


\end{document}